\documentclass[12pt,sort&compress]{article}
\usepackage[dvips]{color}
\usepackage{amsmath}
\usepackage{graphicx} 
\usepackage{tabularx}
\usepackage{cite}
\usepackage{amsfonts}
\usepackage{subcaption}
\usepackage{tabularx,ragged2e}
\usepackage{xcolor}
\newcolumntype{C}{>{\Centering\arraybackslash}X} % centered "X" column
\usepackage[outercaption]{sidecap}    
\usepackage{hyperref}

\newcommand{\gener}{\mathcal{C}}
\newcommand{\Cstat}{\mathcal{C}^{\mathrm{stat}}}
\newcommand{\Dener}{\mathcal{D}}

\begin{document} 
\title{Aggregation with constant kernel under stochastic resetting}
\date{}
\author{Pascal Grange\\
 Department of Physics\\
 Xi'an Jiaotong--Liverpool University\\
111 Ren'ai Rd, 215123 Suzhou, China\\
\normalsize{{\ttfamily{pascal.grange@xjtlu.edu.cn}}}}
\maketitle

\begin{abstract}
 The model of binary aggregation with constant kernel is  subjected to stochastic resetting:
  aggregates of any size explode into monomers at independent stochastic times. These resetting times are 
   Poisson distributed, and the rate of the process is called the resetting rate. 
    The master equation yields a Bernoulli-type equation in the generating function of the concentration of aggregates of any size, which can be solved exactly.
  This resetting prescription leads to a non-equilibrium steady state for the densities 
   of aggregates, which is a function of the size of the aggregate, rescaled by a function of the resetting rate.
   The steady-state density of aggregates of a given size is maximised if the resetting rate is set to the 
    quotient of the aggregation rate by the size of the aggregate (minus one).

\end{abstract}

% large optimal rates and diffusion?

% assumptions

% scaling comme il faut

% comparison with ordinary

% reversibility depolymerisation

\tableofcontents

\section{Introduction}

 Resetting a stochastic process to its initial configuration 
  effectively cuts off long excursions in the space of configurations. In particular,
 the first-passage time of a single diffusive  random walker  is  made finite by resetting the random walker to its initial position at Poisson-distributed stochastic times \cite{evans2011diffusion}. Moreover
  the expectation value of the first passage time at a fixed target  can be optimised as a function of the resetting rate \cite{evans2011optimal}. 
   Optimisation properties of diffusive search times and relaxation dynamics are illustrated in \cite{ghosh2018first,grange2020entropy,grange2020susceptibility}.
    Moreover, stochastic resetting induces non-equilibrium steady states: the steady state of the diffusive random walker with resetting to the origin 
     has been shown to be an exponentially decaying function of the distance to the origin {{\cite{evans2011diffusion}}}. 
   These rich features of stochastic resetting have found numerous applications to
  active matter\cite{evans2018run,refractory},  predator-prey dynamics \cite{mercado2018lotka,toledo2019predator}, population dynamics \cite{da2018interplay,ZRPSS,ZRPResetting}, 
 as well as stochastic processes \cite{lapeyre2019stochastic,gupta2018stochastic,basu2019long,basu2019symmetric,miron2020diffusion,pelizzola2020simple} (see \cite{topical} for a recent review, and references therein).\\
   
  Extensions to many-body interacting systems include reaction-diffusion systems. In particular,
the coagulation-diffusion model under resetting has been studied in \cite{durang2014statistical}. 
On the other hand, in models of aggregation, diffusion or mixing  is supposed to be fast enough so that 
 concentrations are  globally well  defined at all times. Aggregation provides illustrations of  features
of non-equilibrium phenomena, such as steady states and scaling. 
  In the simplest model of aggregation, 
   clusters of all sizes merge pairwise at a uniform rate. This model was solved for the first time by  Smoluchowski  in \cite{von1917mathematical} (see \cite{leyvraz2003scaling,aldous1999deterministic} and  Chapter 5 of \cite{kineticView} for reviews).\\

 {{
   Coupling aggregation to fragmentation gives rise to a very broad family of models, whose  kinetics may be 
    studied to model physical phenomena at various scales, such as phase separation in alloys, nucleation of droplets or the formation of galaxies.
     The convergence properties and structure of equilibrium of the corresponding equations have been studied mathematically (see \cite{wattis2006introduction} for a review).
      Cluster sizes can vary by one unit at a time in the Becker--Doering theory \cite{becker1935kinetische,ball1986becker,jabin2003rate},
      generalisations including all possible processes were presented in \cite{BallCarr}.  }\\

   {{ On the other hand, large systems of particles evolving stochastically can be mapped to a
    random walk in the space of  population sizes. This mapping  has been used to study models of  relaxation with entropy barriers \cite{godreche1995entropy,godreche1996long},
     such as the backgammon model \cite{ritort1995glassiness,backgammon,franz1995dynamical}. The recent applications  of stochastic resetting
      in statistical physics suggest to couple aggregation to fragmentation processes that correspond to resetting the constituents to their original situation 
       as monomers. This fragmentation prescription is the opposite as the one taken in Becker--Doering theory.  
       It is natural to expect  steady states to emerge (together with an exact description). Optimisation properties w.r.t. the resetting 
        rate would generalise to clustering the results obtained for diffusion in \cite{evans2011optimal}.
   In this work we therefore
  subject the Smoluchowski model  to resetting according to a process in which any 
   cluster can explode into monomers at Poisson-distributed times. The resetting rate introduces an additional time scale into the model, and low resetting rates should favour
     large aggregate sizes. Moreover, it is natural to ask whether the scaling properties of the Smoluchowski model (in which large aggregate sizes are scaled 
      by a function of time) are reflected in the steady state of the system under resetting (with sizes rescaled by some function of the resetting rate). }}\\
  
  %one-parameter family

  We will make the same assumptions  as in the Smoluchowski model: the kinetics of the reactions 
   does not depend on the shape of the aggregates, and the transport phenomena are fast enough 
    for the concentration of aggregates of any size to be a well-defined function of time. With these assumptions,
     the concentrations  evolve according to a set of coupled  master equations. These master equations  induce a non-linear equation
      in the generating function of concentrations.\\
     
     In Section 2 we  set the notations and work out the master equation induced by the resetting prescription.
      In Section 3 the total density of  clusters is expressed as a function of time, which allows to solve the
       master equation as a Bernoulli equation. 
        In Section 4 the stationary state is studied: in particular, the 
        concentration of aggregates of all masses are expressed, and  maximised in  the resetting rate. {{The limit of low resetting rate is shown to yield the Smoluchowski model}}. 
     In Section 5 initial conditions consisting of aggregates of uniform  size are studied. To obtain an idea of the typical size
      of aggregates,  the second moment of the density is expressed as a function of time.

\section{Model and quantities of interest}

 Consider the aggregation process of identical monomers with constant kernel. Each of the processes in which a cluster of size $i$  (denoted by $A_i$)
   and a cluster of size $j$  join to form a cluster of size $i+j$, is described by a reaction
\begin{equation}\label{aggregationProcess}
 A_i + A_j \longrightarrow A_{i+j}
\end{equation}
 of rate $K>0$, independent of the size (and shape) of the clusters. Let us introduce reversibility into the process under the form of resetting.
  In an infinitesimal interval $d\tau$ of time, any aggregate of size $k$ has a probability $\rho d\tau$ of exploding into 
   $k$ monomers in the reaction
   \begin{equation}\label{resettingProcess}
 A_k  \longrightarrow k A_1.
\end{equation}

 Let us rescale time so that the rate of aggregation equals $2$. The rescaled time is denoted by $t$. The resetting rate is denoted by $r$ in the rescaled time: 
 \begin{equation}\label{rescaledTime}
 t :=  K \frac{\tau}{2},\;\;\;\;\;\;\;\;\;\;\;\;\; r = \rho \times \frac{\tau}{t} = \frac{2\rho}{K}.
\end{equation}
 The main quantities of interest are the  concentrations of aggregates of all sizes:
\begin{equation}
 \left\{c_k( t ) :=  \mathrm{concentration}\; \mathrm{of}\;\mathrm{aggregates}\; \mathrm{of}\; \mathrm{size}\; k \;\mathrm{at}\; \mathrm{time}\; t,\;\;k\geq 1,\;\;t>0 \right\}.
\end{equation}
 The aggregates are assumed to be well mixed in a solvent, so that the above densities are well defined at all times, and 
  the monomers resulting  from the resetting processes of Eq. (\ref{resettingProcess}) are immediately  available for aggregation.\\

Consider the concentration of aggregates of size $k$, for some $k\geq 1$. It satisfies the following master equation:
\begin{equation}\label{masterck}
\frac{dc_k}{d{{t}}} = \sum_{i+j = k}  c_i c_j - c_k \sum_i  c_i - r  c_k + r \delta_{k1}\sum_{i}    i  c_i.
\end{equation}
The first two terms on the r.h.s. correspond to aggregation of pairs of clusters (of sizes $i$ and $j$)  into one cluster of size $k$, the second one to the 
  aggregation of a cluster of size $k$ and another cluster of any size $i$. These two terms are those present in the irreversible model \cite{von1917mathematical}.
   The third term corresponds to the resetting of an aggregate of size $k$ to $k$ monomers at rate $r$, and the last term expresses  the contribution of the
    resulting monomers to the concentration $c_1$. {{As the dissociation of aggregates into monomers is the only dissociation channel we consider, this  term can be added to the master equation for all values of $k$, with a factor of $\delta_{k1}$}}. For $k=1$, the contribution of the resetting processing from $c_1$ to the time derivative reads $(-rc_1 + r \times 1 \times c_1)=0$, which is consistent  because the resetting of an aggregate of size $1$ leaves it unchanged.\\

  Let us denote by $\gener$ the generating function of the densities of aggregates,
   and by  $N$ the total density of aggregates:
  \begin{equation}
  \begin{split}
  \gener(t,z) &:= \sum_{k\geq 1 } c_k(t) z^k,\\
  N(t) &:= \sum_{k\geq 1} c_k(t) = \gener( t, 1 ).
  \end{split}
 \end{equation}
  {{The  generating function of  the densities not only  allows to read off the coefficients $(c_k(t))_{k\geq 1}$ from a series expansion, but it also gives access to the moments of the distribution of densities, by taking derivatives w.r.t. the variable $z$ at the value $z=1$. These moments give physical insights into the distribution of mass (for instance the ratio of  the moment of order $1$ to the moment of order zero, $N(t)^{-1}\sum_k k c_k(t)$ is the average mass of aggregates at time $t$).}}\\

  For our purposes it is enough to restrict the variable $z$ to $[0,1]$.  Both aggregation and resetting (Eqs \ref{aggregationProcess},\ref{resettingProcess}) conserve mass.
 The total mass density is therefore a constant, as in the model without resetting. Let us denote it by $M$:
 \begin{equation}
  M := \sum_{k \geq 1} k c_k(t).
 \end{equation}
  The master equations  for aggregates of fixed size (Eq. (\ref{masterck})) induce the following master equation
   for the generating function:
   \begin{equation}\label{evolC}
 \frac{\partial \gener(t,z)}{\partial t} =  \gener(t,z)^2 - (2N(t) + r)\gener(t,z) + rMz.
 \end{equation} 
  Setting the total mass density $M$ to unity is equivalent to picking a unit of volume, just as setting the
   rate  of aggregation to $2$ is equivalent to rescaling time:
   \begin{equation}\label{setM}
    M:=1.
   \end{equation}

 {\bf{Monomer-only initial conditions.}} All the equations so far are independent of the initial conditions.
 For definiteness we can consider the monomer-only  initial conditions, where all the aggregates have unit size, with a unit total mass density:
\begin{equation}\label{pure}
 c_k(0) = \delta_{k1}.
\end{equation}
 We will use these boundary conditions in Section 3.2, but eventually we will consider more general initial configurations of densities.

%With the monomer-only initial conditions, 
% \begin{equation}
 % N(0) = c_1(0) = 1,\;\;\;\;\;\;\;\; M=1.
  %\end{equation}
  
%  Let us set all the entries of the aggregation kernel to a constant (setting the time scale of the process), and the resetting 
% rates to a uniform resetting rate denoted by $r$:
% \begin{equation}
% \forall i,j \geq 1,\;\;\;\;\;\;\; K_{ij} := 2,\;\;\;\;\;\;\;\;   L_i:=r.
% \end{equation} 

 \section{Solution of the master equation}  
 \subsection{Total density of clusters}
 The evolution equation for the total density of clusters is obtained by substituting $1$ to $z$ in Eq. (\ref{evolC}):
  \begin{equation}\label{evolN}
 \frac{ d N}{d t} =  -N^2 - rN + r,
 \end{equation}  
 where we used the value of the mass concentration defined in Eq. (\ref{setM}). The r.h.s. is quadratic in $N$, the two roots have opposite signs, let us 
  denote by $N_-$ and $N_+$.
 The long-time limit $N(\infty)$ of the total density is equal to the positive root:
  \begin{equation}
 N(\infty) = N_+ := \frac{r}{2}\left(-1 + \sqrt{1 + \frac{4}{r}}\right),\;\;\;\;\;\;\; N_- := \frac{r}{2}\left(-1 - \sqrt{1 + \frac{4}{r}}\right).
 \end{equation} 
 The value of $N(\infty)$  depends on the resetting rate. 
   At low resetting rate  it is close to zero (equivalent to $\sqrt{r}$), and at large resetting rate  it is close to  $1$ (it grows towards $1$ when the resetting rate goes to infinity, in which limit the resetting process destroys the aggregation process).\\

 %CODE: figure_N
 \begin{figure}
\includegraphics[width=18cm]{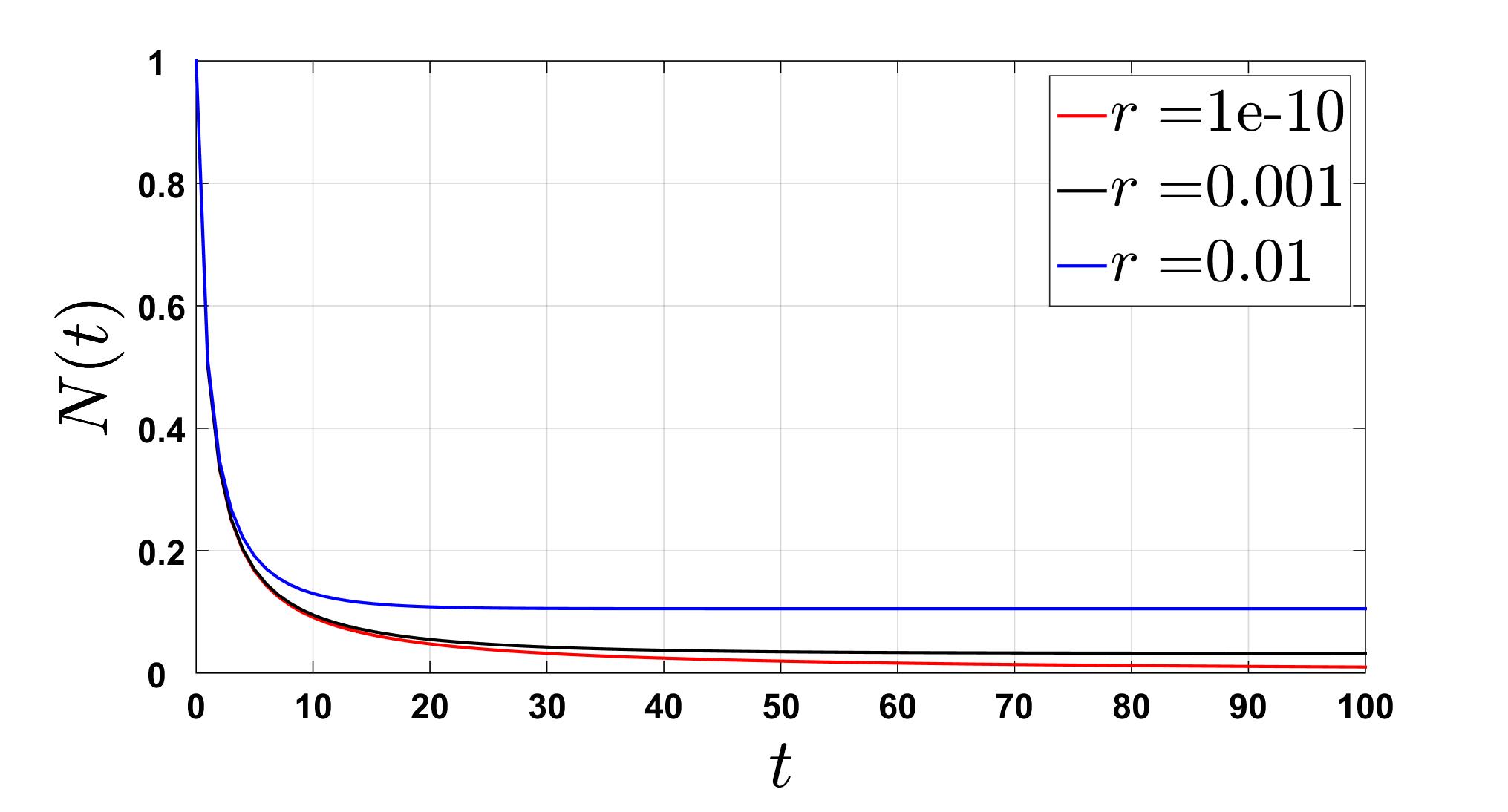} 
\caption{{\bf{Total density of aggregates as a function of time}}. The plot corresponds to monomer-only initial conditions ($N(0)=1$). The large-time limit is a growing function of the resetting rate.}
\label{figureN}
\end{figure}

%CODE: figure_ck
%figure_ck_log

% \begin{figure}
%\includegraphics[width=18cm]{figureck.jpg} 
%\caption{{\bf{Density of aggregates $c_k$  (of fixed size $k$) as a function of the resetting rate $r$.} For better visibility each of the plot is normalised by its maximum 
% value.}}
%\label{figureck}
%\end{figure}

\begin{figure}
\begin{subfigure}{\textwidth}
  \centering
  \includegraphics[width=1\linewidth]{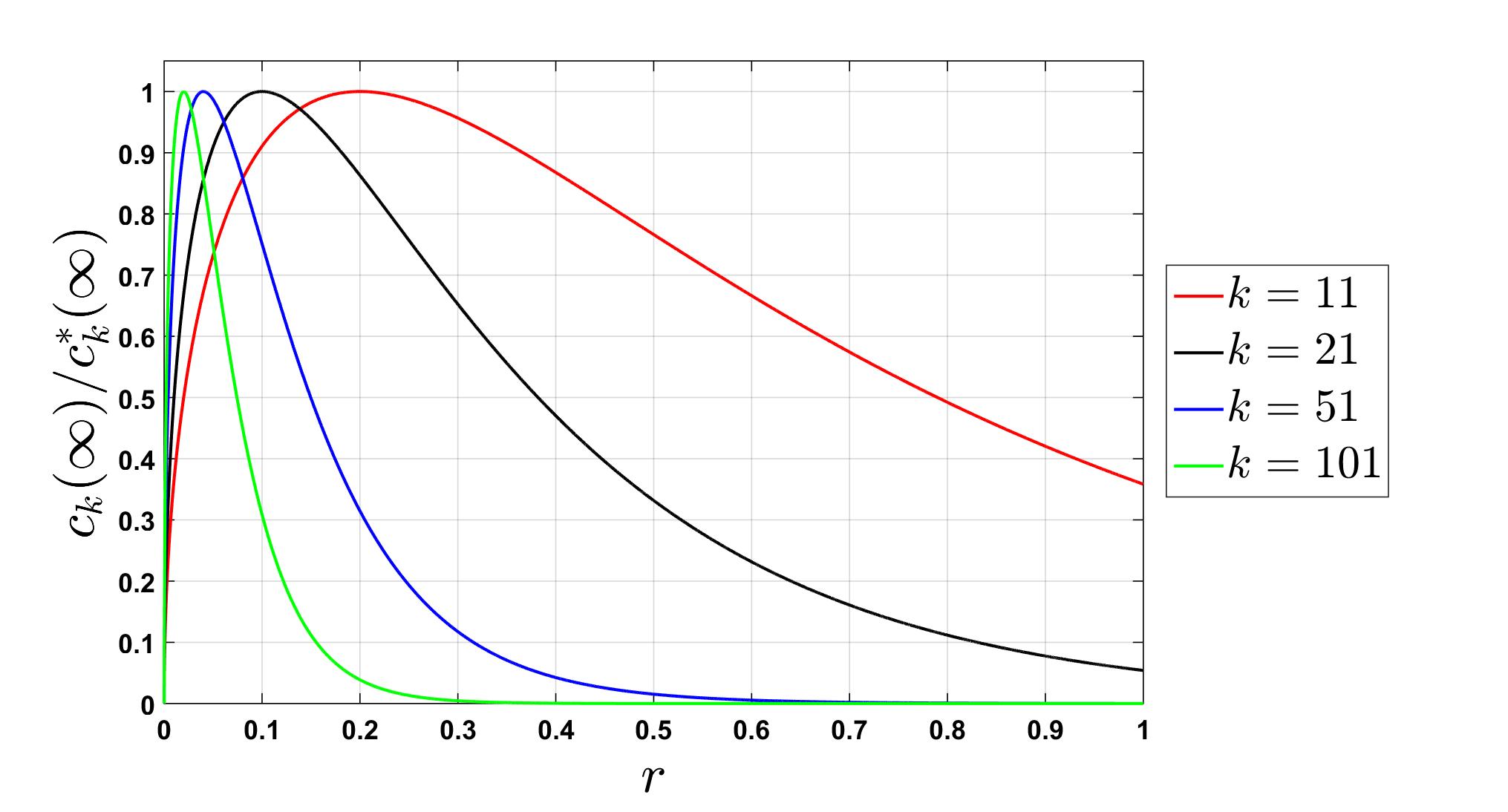}  
  \caption{}
  \label{fig:sub-first}
\end{subfigure}
\newline
\begin{subfigure}{\textwidth}
  \centering
  \includegraphics[width=1\linewidth]{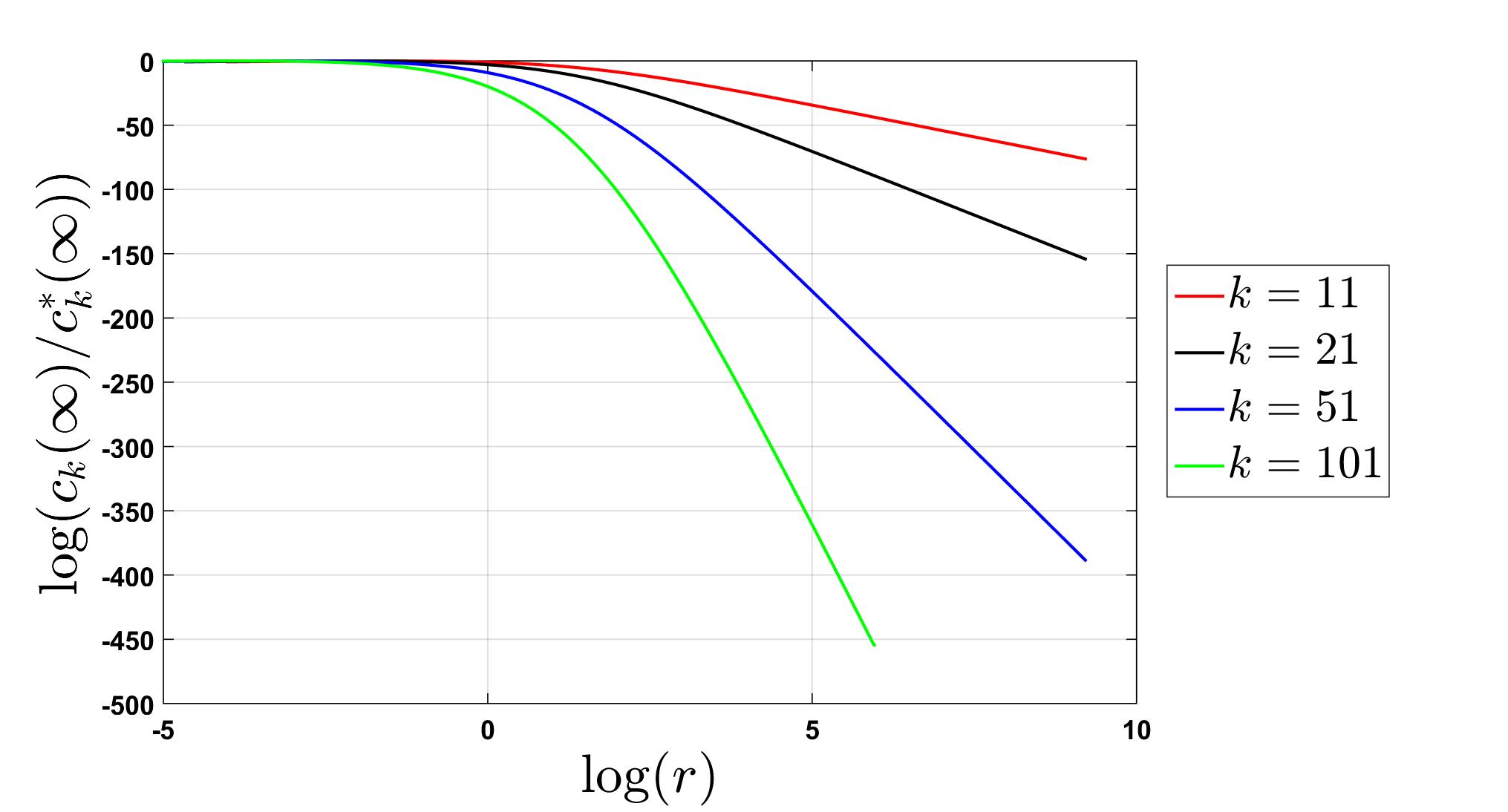}  
 \caption{}
  \label{fig:sub-third}
\end{subfigure}
\caption{{\bf{Stationary density of aggregates $c_k$  (of fixed size $k$) as a function of the resetting rate $r$.}} (a) The maximum of each plot is at $r^\ast_k = 2/(k-1)$. (b) A log-log plot shows that the steady-state density at large resetting rate decays like a power of $r$.}
\label{figureck}
\end{figure}

%CODE: figure_ck_optimal
\begin{figure}
\includegraphics[width=18cm]{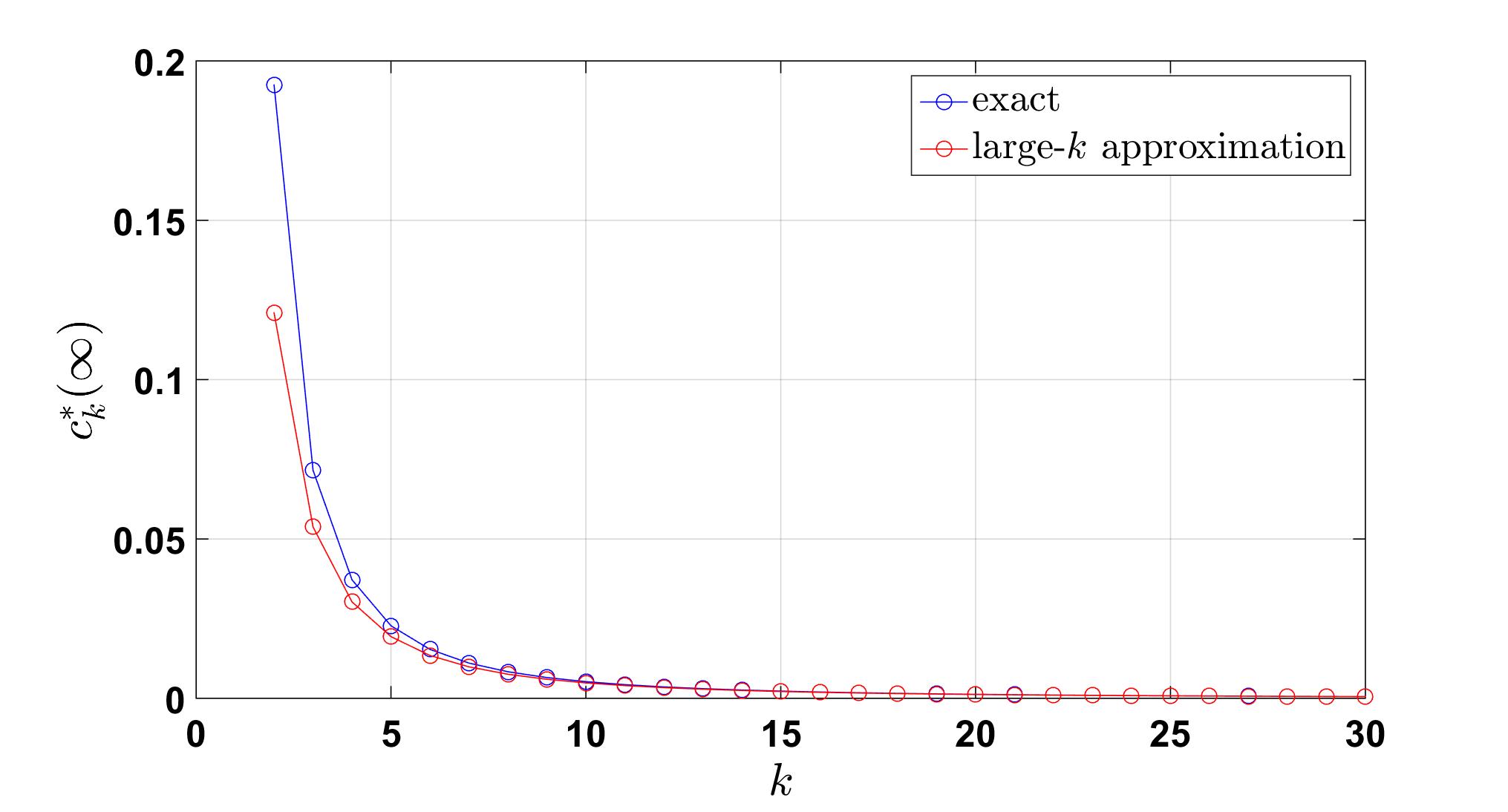} 
\caption{{\bf{Optimal density of aggregates $c_k( r^{\ast}_k)$   as a function of the size $k$ of the aggregate.}} The large-$k$ approximation is given by $\sqrt{2/(e\pi)}k^{-2}$.}
\label{figureckOptimal}
\end{figure}

 With these notations we can rewrite Eq. (\ref{evolN}) as
  \begin{equation}
  -1 = \frac{1}{(N(t) - N_-)(N(t) - N_+)} \frac{ d N}{d t} = \frac{1}{N_+ - N_- }\left( -\frac{1}{N-N_-}  + \frac{1}{N- N_+} \right) \frac{ d N}{d t},
 \end{equation} 
  \begin{equation}\label{toInt}
 - (N_+ - N_-) =  \frac{d}{dt}\left(- \log|N(t) - N_-| + \log| N(t) - N_+| \right).
 \end{equation} 
  {{The aggregation process tends to decrease the total density of aggregates, while the resetting process increases it by 
   turning large clusters into monomers. If the total density of aggregates starts with $N(0)>N_+$ (as in the monomer-only initial conditions), 
    aggregation drives the density of aggregates towards lower values at the beginning of the process  (in the case of monomer-only boundary conditions,
     aggregation is the only phenomenon that can take place at the beginning of the process). If the total density of clusters starts from below $N_+$, resetting  drives the total density of clusters towards lower values at the beginning of the process (indeed, in the case of initial conditions with only polymers of large size, aggregation decreases the total density at a slow rate, compared to the rate of  increase contributed by the decay of  large polymers into monomers).}}\\
     
     {{
      Moreover, the time-derivative of  the total density 
      can only be zero if the density equals  $N_+$ (because $N_-$ is negative). If the process starts at $N(0)>N_+$, the density cannot go below $N_+$ because it would need to reach a minimum before approaching  the asymptotic value $N_+$. If the process starts at $N(0)<N_+$, the density cannot go above $N_+$ because it would need to reach a maximum. The quantity $N(t) - N_+$ therefore has constant  sign. Moreover the quantities $N(t)-N_-$ and $N(0)-N_-$ are both positive because $N_-$ is negative.}}  
 Integrating Eq. (\ref{toInt}) between time $0$ and time $t$ therefore yields:
  \begin{equation}
 - \sqrt{r(r+4)} t = \log \left(  \frac{(N(t) - N_+)(N(0)-N_-) }{(N(t) - N_-)(N(0) - N_+) } \right),
 \end{equation}
 from which obtain the exponential convergence of the total density of cluster to the steady state value:
% \begin{equation}
% N(t) = \frac{-r + \sqrt{r(r+4)}}{2} - \frac{\sqrt{r(r+4)}}{\frac{2+r +  \sqrt{r(r+4)}}{-2-r +  \sqrt{r(r+4)}}e^{ \sqrt{r(r+4)}t} +1}.
% \end{equation}
% 
 \begin{equation}\label{N(t)}
  N(t) = \frac{N_+(  N(0 ) - N_-) {{-}} N_- (N(0) - N_+ )e^{-\sqrt{r(r+4)}t}}{  N(0 ) - N_-  - ( N(0) - N_+  ) e^{-\sqrt{r(r+4)}t}  }.
 \end{equation}
  The total density obtained in Eq. (\ref{N(t)}) holds for any initial distribution of cluster sizes with a unit mass concentration. The dependence on the 
  initial condition is entirely contained in the initial density of clusters $N(0)$. {{This result is plotted on Fig. \ref{figureN} for monomer-only initial conditions}.}\\

\subsection{Generating function for monomer-only initial conditions}
 
 Going back to Eq. \ref{evolC} and introducing the new function
 \begin{equation}\label{defDener}
  \Dener(t,z) = \gener(t,z) - N(t),
 \end{equation}
  we obtain a non-linear differential equation in time:
   \begin{equation}\label{evolD}
   \frac{\partial \Dener}{\partial t} =  \Dener^2 - r \Dener  + r( z - 1 ).
 \end{equation}
   The last term does not depend on the variable $t$ and can therefore be treated as a constant. Let us denote by
    $X(z)$ a root (we will specify which one later)  of the quadratic equation on the r.h.s.: 
    \begin{equation}\label{eqx}
     X(z)^2 - r X(z) + r( z-1) = 0. 
    \end{equation}
 We can convert Eq. (\ref{evolD}) into a Bernoulli equation by changing unknown from $\Dener$ to $F$ as follows:
\begin{equation}\label{defF}
   \Dener(t,z) =: F(t,z) + X(z),
    \end{equation}
  \begin{equation}\label{evolF}
   \frac{\partial F}{\partial t} =  F^2 + (2X - r) F.
 \end{equation}
 Changing function again through the definition
  \begin{equation}\label{defG}
   G(t,z) := \frac{1}{F(z,t)}
 \end{equation}
 and dividing Eq. (\ref{evolF}) by $F^2$ yields 
 \begin{equation}
   \frac{\partial G}{\partial t}(t,z) = (r - 2X(z)) G(t,z) - 1.
 \end{equation}
  Solving this first-order ODE in $t$ involves a $z$-dependent  integration constant, denoted by $Y(z)$, such that 
  \begin{equation}\label{exprG}
   G(t,z) = Y(z) e^{(r-2X(z))t} + \frac{1}{r - 2X(z)}.
 \end{equation}
  
  The generating function is therefore expressed (using Eqs (\ref{defG},\ref{defF},\ref{defDener})) as
   \begin{equation}\label{solC}
   \gener(t,z) = N( t ) + X( z ) + \frac{r - 2X(z)}{( r -2X(z))Y(z) e^{(r-2X(z))t } + 1 }.
 \end{equation}
  For the generating function to have a finite limit  at large time, we must  pick the negative root of Eq. (\ref{eqx}):
   \begin{equation}\label{XDef}
   X( z ) := \frac{1}{2}\left( r  - \sqrt{r^2 + 4r(1-z)}\right).
   \end{equation}
  We notice that $X(z)$ does not depend on the choice of initial conditions.  
   The integration $Y(z)$ constant introduced in Eq. (\ref{exprG}) can be traded for:
 \begin{equation}\label{solK}
   K(z):= ( r -2X(z))Y(z).
 \end{equation}
    Imposing the monomer-only  initial condition of Eq. (\ref{pure}) yields
    \begin{equation}\label{initCondz}
    \begin{split}
 &\forall z \in [0,1],\;\;\;\;   \gener(0,z) = z,\\
   &{\mathrm{hence}}\;\;\;  z =  1+ X( z ) + \frac{r - 2X(z)}{K(z) + 1 }.
   \end{split}
 \end{equation}
 Hence, using Eq. (\ref{XDef}), we obtain
 \begin{equation}\label{KDef}
   K( z ) = -1 + \frac{2  \sqrt{r^2 + 4r(1-z)}}{ 2\left(z-1\right) - r + \sqrt{r^2 + 4r(1-z)} }.
 \end{equation}
 We notice that $|K(z)|$ goes to infinity when $z$ goes to $1$.
 Moreover, at all times $C(t,1) = N(t)$, hence Eq. (\ref{solC}) reduces to
  \begin{equation}
    X( 1 ) + \frac{r - 2X(1)}{K(1) e^{(r-2X(1))t } + 1 } = 0,
 \end{equation}
 which is consistent since $X(1)=0$ and $|K(1)|=\infty$.\\
% Moreover, 
%  \begin{equation}
%   r- 2 X(z ) = +\sqrt{r^2 + 4r( 1 -z)}.
%  \end{equation}

   Rearranging  Eqs (\ref{solK},\ref{solC}) we obtain the generating function of concentrations for monomer-only initial conditions:
   \begin{equation}\label{solGener}
    \gener(t,z) = N( t ) + \frac{1}{2}\left( r  - \sqrt{r^2 + 4r(1-z)}\right) + \frac{ \sqrt{r^2 + 4r(1-z)} e^{-  \sqrt{r^2 + 4r(1-z)} t}}{ -1 + \frac{2 \sqrt{r^2 + 4r(1-z)}}{ 2(z-1) - r +  \sqrt{r^2 + 4r(1-z)}}   + e^{ -  \sqrt{r^2 + 4r(1-z)}   t}}.
   \end{equation}

    \section{Stationary state}  
  \subsection{Stationary density profile as a function of the resetting rate}
 
   The large-time limit  of the generating function reads
     \begin{equation}\label{generIn}
    \gener(\infty,z) = N( \infty ) + X(z)=\frac{r}{2} \sqrt{1 + \frac{4}{r}}-  \frac{1}{2} \sqrt{r^2 + 4r(1-z)}.
   \end{equation}
    As a check, we can solve directly  the  equation 
  satisfied by the steady state $\Cstat$, which is obtained by putting time-derivatives to zero in
  the master equation (Eq. (\ref{evolC})):
    \begin{equation}
 \Cstat(z)^2 - (2N(\infty) + r)\Cstat(z) + rz = 0.
 \end{equation} 
  Using the fact that  $\Cstat(0) = 0$   selects the solution
   \begin{equation}\label{Cstat}
 \Cstat(z) = \frac{\sqrt{r(r+4)}}{2}\left( 1 - \sqrt{1 - \frac{4z}{r+4}} \right),
 \end{equation} 
 which is indeed equal to the large-time limit of the generating function  $\gener(\infty,z)$, obtained in Eq. \ref{generIn}.
  Moreover, this  state is independent of the initial conditions, because $N(\infty)$ and $X(z)$ are. At large time the system forgets 
   its initial conditions, because they only enter the expression of the generating function through the quantity we denoted by $Y(z)$ 
    in Eq. (\ref{solC}), whose contribution is exponentially suppressed at large time.\\
 
   Expanding in powers of $z$ yields the expression of the steady-state density $c_k(\infty)$ of the 
    clusters of size $k$. Indeed, using $\Gamma(1/2) = \sqrt{\pi}$, we may  substitute $4z/(r+4)$ to $s$ in the expansion
    \begin{equation}
    \sqrt{1-s} = 1- \sum_{k\geq 1}\frac{\Gamma\left( k - \frac{1}{2}\right)}{\sqrt{\pi} \Gamma( k + 1 )}s^k,
    \end{equation}
    to obtain
    \begin{equation}\label{ckstat}
    \begin{split}
 &\Cstat(z) = \sum_{k\geq 1}  c_k(\infty) z^k,\\
  &c_k(\infty) = \sqrt{r(r+4)}\frac{\Gamma\left( k - \frac{1}{2}\right)}{2\sqrt{\pi} \Gamma( k + 1 )}\left( 1 + \frac{r}{4}\right)^{-k}.
  \end{split}
 \end{equation}

     %{{\subsection{Scaling form of the concentrations of large aggregates}}}
     {{
      In the original time denoted by $\tau$,
     consider a rescaling of the rate of aggregation $K$ of aggregation by a factor of $\alpha$ (at fixed resetting rate $\rho$, in the notations of Eq. (\ref{rescaledTime})).  
      Because of the redfinition  
       of time, we are led to the same equations of motion as before, with primed symbols 
        for time and resetting rate  $t'$ and $r'$ substituted to $t$ and $r$ respectively:
        \begin{equation}
        K':=\alpha K, \;\;\;\; t' = K'\frac{\tau}{2} = \alpha t,\;\;\;\;\;\;\; r' = \rho \times \frac{\tau}{t'} = \frac{2\rho}{K'} = \frac{1}{\alpha K}.
        \end{equation}
       The effect on the equation of motion is therefore identical to a rescaling of the rate $r$ by $\alpha^{-1}$. For this value of $r'$, we can read off
        the steady-state concentrations from Eq. (\ref{concProfile}). The large-time behaviour of the model is therefore captured by the single parameter $r$, 
         instead of the pair $(K,\rho)$. We can therefore reason on the large sizes using the resetting rate only, at fixed aggregation rate.
       Intuitively, low resetting rates favour large values of $k$, so we can look for a characteristic size $\sigma(r)$, where $\sigma$ is a decreasing  
        function of $r$, such that the stationary concentration of large aggregates assumes the scaling form
        \begin{equation}
         c_k(\infty) \underset{k\to\infty}{\sim} \nu(r) g\left(\frac{k}{\sigma(r)} \right),
        \end{equation}
    where $g$ is a scaling function to be determined, and the prefactor  $\nu$  ensures the conservation of mass, $\sum_{k} kc_k(\infty) = 1$.\\}}

       {{
     Inspection of the exact result of Eq. (\ref{ckstat}) proves that we only need an equivalent at large $k$  of the quotient of the two values of the Gamma
     function to read off the scaling form.}}
      Using the equivalent 
    \begin{equation}\label{equivGamma}
    \Gamma(  k -1/2)  \underset{k\to\infty}{\sim} \Gamma(k+1) k^{-3/2}
    \end{equation}
     yields the following 
     equivalent for the stationary concentration of aggregates of large size:
     \begin{equation}\label{concProfile}
      c_k(\infty) \underset{k\to\infty}{\sim} \frac{\sqrt{r(r+4)}}{2\sqrt{\pi} }k^{-\frac{3}{2}} \left( \frac{4}{r+4} \right)^k.
     \end{equation}

     The concentration $c_k(\infty)$  therefore assumes a scaling form, 
      with the gamma distribution of parameters $-1/2$ and 1:
    \begin{equation}
     \begin{split}
      c_k(\infty) &\underset{k\to\infty}{\sim}  \frac{\sqrt{r(r+4)}}{2\sqrt{\pi}}\left( \log\left( 1 + \frac{r}{4}\right) \right)^{\frac{3}{2}}g\left( \frac{k}{\sigma(r)} \right),\\
      {\mathrm{with}}\;\;\;\; g(x)& := x^{-\frac{3}{2}}e^{-x},\;\;\;\;\;{\mathrm{and}}\;\;\;\sigma(r) := \left(\log\left( 1 + \frac{r}{4}\right) \right)^{-1}.
      \end{split}
     \end{equation}

   \subsection{Size-dependent optimal resetting rate}
    For any  value $k$  of the cluster size, the steady-state density depends on the resetting rate through the function
    \begin{equation}
     \varphi_k(r) :=\sqrt{r(r+4)}\left( \frac{4}{r+4} \right)^k.
    \end{equation}     
  For any size $k>1$, the steady-state density therefore goes to zero in the limit of large resetting (and it goes to zero as $2\sqrt{r}$ in the limit of 
    small $r$). There is therefore a value 
     of the resetting rate  that maximises the steady-state density at cluster size $k$ (except for $k=1$, as the density of monomers is maximised in the limit of infinite resetting rate, which destroys aggregation altogether). 
     Calculating the derivative of $\varphi_k$ yields the unique optimal value $r_k^\ast$ of the  resetting rate:
      \begin{equation}
     \frac{1}{2 r_k^\ast} = \left( k {{-}} \frac{1}{2}\right)\frac{1}{ r_k^\ast +4},
      \end{equation}
     \begin{equation}
     r_k^\ast = \frac{2}{k{{-1}}},\;\;\;\;\;\;\;\;\;{\mathrm{for}}\;\;\;k>1.
      \end{equation}
       {{Substituting this optimal value to the resetting rate in Eq. (\ref{ckstat})}}, the maximum value of the density of aggregates of size $k$ therefore reads
%       \begin{equation}
%       {{c_k^\ast(  \infty ) = \frac{\sqrt{(2k+1)}}{k} \frac{\Gamma\left( k - \frac{1}{2}\right)}{\sqrt{\pi} \Gamma( k + 1 )} \left(1 + \frac{1}{2k} \right)^{-k}. }}
%       \end{equation}
       \begin{equation}\label{toExpand}
       {{c_k^\ast(  \infty ) = \frac{\Gamma\left( k - \frac{1}{2}\right)}{\sqrt{\pi} \Gamma( k + 1 )} \sqrt{\frac{2}{k-1}}\left(1 + \frac{1}{2(k-1)} \right)^{-k}. }}
       \end{equation}
        The optimal value of resetting  goes to zero at large sizes, which is intuitive as 
       rare resetting events favour the formation of large aggregates. {{The steady-state density (normalised by the maximum $c_k^\ast(  \infty ) $) is plotted for a few values of  $k$ on Fig. (\ref{figureck}(a)). At fixed resetting rate, the steady-state density goes to zero in the limit of large size, because large clusters are penalised by the resetting process. Moreover, the density goes to zero at large resetting rate and fixed size as $r^{-(k-1)}$, which is illustrated by the log-log plot on Fig. (\ref{figureck}(b)).}}\\

        Moreover, Eq. (\ref{equivGamma}) yields the large-$k$ equivalent of the optimal value 
               \begin{equation}\label{optimEquiv}
        c_k^\ast(   \infty )  \underset{k\to\infty}{\sim}\sqrt{ \frac{2}{\pi e}} \frac{1}{k^2}.
       \end{equation}
       {{ The optimal values of the densities are plotted as a function of the size of the aggregate of Fig. (\ref{figureckOptimal}), where 
        the large-size equivalent is also plotted. To estimate how fast  the large-size regime is reached,  we can work out 
         the leading correction to the asymptotic behaviour. We need the next term in Eq. (\ref{equivGamma}). Let us start  from the  
     following  large-$m$  asymptotic expansion \cite{dyson2013lehmer}:
       \begin{equation}\label{useful}
        \frac{\Gamma\left( m + \frac{1}{2} \right) }{\Gamma( m  )}= \sqrt{m}\left( 1- \frac{1}{8m} + O\left( m^{-2}\right)\right),
       \end{equation}
     from which we can work out a  correction of order $k^{-1}$ to Eq.  (\ref{optimEquiv}). Indeed, we need the large-$k$ behaviour of
     \begin{equation}
     \begin{split}
     \frac{\Gamma\left( k - \frac{1}{2}\right)}{\Gamma( k + 1 )} &= \frac{\Gamma\left( k - \frac{1}{2}\right)}{k(k-1)\Gamma(k-1)}\\
     &= \frac{1}{k^2}\frac{1}{1-\frac{1}{k}}\sqrt{k-1}\left(    1- \frac{1}{8k} + O\left( k^{-2}\right)\right) \\
     & = k^{-3/2}\left( 1+\frac{1}{k} + O(k^{-2})\right) \left( 1-\frac{1}{2k} + O(k^{-2})\right) \left(    1- \frac{1}{8k} + O\left( k^{-2}\right)\right) \\
     & =  k^{-3/2}\left( 1+\frac{3}{8k} + O(k^{-2})\right),\\
     \end{split}
     \end{equation}
     where in the second line we used Eq. (\ref{useful}) with $m=k-1$.\\}}
    
     {{
     Another correction of order $k^{-1}$ comes from
     \begin{equation}
     \begin{split}
      \left(1 + \frac{1}{2(k-1)} \right)^{-k} &= \exp\left( -k \log\left( 1 + \frac{1}{2(k-1)}\right)    \right) \\
         &= \exp\left( -k \log\left( 1 + \frac{1}{2k} + \frac{1}{2k^2} + O( k^{-3}) \right)\right)\\
        & = \exp\left( -k \left( \frac{1}{2k} + \frac{1}{2k^2}  -  \frac{1}{8k^2} + O( k^{-3}) \right) \right)\\
         &= \frac{1}{\sqrt{e}}\left(   1-\frac{3}{8k} + O( k^{-2}) \right).
      \end{split}
     \end{equation}
     Going back to Eq. (\ref{toExpand}), we collect the leading corrections as
     \begin{equation}
         c_k^\ast(   \infty ) =\sqrt{ \frac{2}{\pi e}} \frac{1}{k^2} \left(1 + \frac{1}{2k} +  O( k^{-2}) \right).
     \end{equation} 
     The asymptotic value of Eq. (\ref{optimEquiv}) is therefore approached from above in the limit of large size.
   }}
       
%    The following asymptotic series is occasionally useful in probability theory (e.g., the one-dimensional random walk):
%
% (Gamma(J+1/2))/(Gamma(J))=sqrt(J)(1-1/(8J)+1/(128J^2)+5/(1024J^3)-(21)/(32768J^4)+...) 	
%(98)
%(OEIS A143503 and A061549; Graham et al. 1994). This series also gives a nice asymptotic generalization of Stirling numbers of the first kind to fractional values.
%   

 \subsection{{Low resetting rate and consistency with the Smoluchowski model}} 
 
 {{The large-time limit of our model presents a non-zero total density, whereas the 
  total density in the Smoluchowski model goes to zero at large time. Indeed, denoting the ordinary system without resetting
   with the symbol $(r=0)$ in an exponent, we have in the case of monomer-only initial conditions
  \begin{equation}\label{SmolN}
   N^{(r=0)}(t) = \frac{1}{1+t},
  \end{equation}
   which is the solution of Eq. (\ref{evolN}) for $r=0$. Of course the master equation of the Smoluchowski model of aggregation is recovered if we set the resetting rate 
    to zero in our master equation (Eq. (\ref{evolC})), but working out the low resetting-rate limit of  our results should yield the densities predicted by the
     Smoluchowski model.}}\\

     {{Let us first address the total density of aggregates. 
     Of course, the asymptotic value $N_+$ goes to zero when the resetting   rate $r$  goes to zero, but the time-dependence displayed in Eq. (\ref{SmolN})
      should be recovered at low resetting in a transitory regime. We can define this regime by fixing time, and looking for a range of resetting rates 
       that ensures that the exponential terms in the expression of the total density are close to $1$. 
        At fixed time $t$, the density expressed in Eq. (\ref{N(t)}) is in a transitory regime if $r$ is low enough for the quantity $\sqrt{r(4+r)}t$ to be close
         to zero. At low $r$, the quantity is equivalent to $2\sqrt{r}t$, and the low-resetting regime is defined
          by 
          \begin{equation}
           r\ll\frac{1}{4t^2}.
          \end{equation}
          In this limit, $N_+\simeq \sqrt{r}$ and $N_-\simeq -\sqrt{r}$.
        Equivalently, we can reason at fixed resetting rate, and declare that  times much smaller than $(2\sqrt{r})^{-1}$  constitute 
         a transitory regime, in which the system is in the situation of the ordinary model of aggregation. Both reasonings lead to the approximation
         \begin{equation}
         N(t)  \simeq \frac{N_+ - N_- }{N_+ - N_- + (1-N_-)2\sqrt{r} t }.
         \end{equation}
        Indeed, at fixed $t$ and low $r$, the term $N_-(1-N_+)\sqrt{r}t$ (present in the numerator of Eq. (\ref{N(t)})) is equivalent to $2rt$, and it therefore subdominant, as all the other terms in the fraction are of order $\sqrt{r}$.
         Keeping only the dominant terms  yields a low-resetting time-independent limit
          \begin{equation}
         N(t)  \simeq \frac{2\sqrt{r}}{2\sqrt{r}+ 2\sqrt{r} t } = \frac{1}{1+t},
         \end{equation}
           which as expected coincides with $N^{(r=0)}(t)$.\\}}

      {{
      The same approach can be taken  to study  the generating function at low resetting rate. 
      Let us fix some time $t$, and some $z$ in $[0,1]$. The resetting rate $r$ is deemed small for these particular values if the exponential relaxation of the 
       generating function has not taken place yet. This yields the condition
    \begin{equation}
     2\sqrt{(1-z)r} t \ll 1,\;\;\;\;\;\;\;\; {\mathrm{i.e.}}\;\;\; r\ll\frac{1}{4(1-z)t^2}.
     \end{equation}   
    In this limit,  the combination $r-\sqrt{r^2+4r(1-z)}$  is equivalent  to $2\sqrt{r(1-z)}$.  
     Both the numerator and the denominator in the last term  in Eq.  (\ref{solGener}) 
      are of order $\sqrt{r}$ (where the exponential factor contributes a factor of $1$ in the numerator and the term $2\sqrt{r(1-z)} t$ by Taylor expansion in the denominator), so their quotient has a finite limit:
     \begin{equation}
     \begin{split}
     C(t,z) &\simeq  N^{(r=0)}(t) + \frac{ 2\sqrt{r(1-z)}}{ \frac{4\sqrt{r(1-z)}}{2(z-1)} - 2\sqrt{r(1-z)} t }\\
       &=  \frac{1}{1+t} + \frac{1}{(z-1)^{-1}-t}\\
       &= \frac{1}{1+t}\left( \frac{z}{1-(z-1)t}\right),  
     \end{split}
     \end{equation}
    which is indeed the generating function of the Smoluchowski model, which satisfies the master equation at $r=0$.\\}}

 \section{Typical size of aggregates for polymer-only boundary conditions} 
   Consider slightly more general initial conditions in which the total mass density $M=1$ results from polymers
    of fixed size $A>1$:
    \begin{equation}
     c_k(0) = \frac{1}{A}\delta_{kA}.
    \end{equation}
    We keep the same unit of time and volume, so  the master equation is unchanged.
     The total cluster density is still given by Eq. (\ref{N(t)}), with $N(0)=A^{-1}$.
     The only modification in the solution  
  comes from the initial condition on the generating function (still denoted by $\gener$):
    \begin{equation}
    \gener(0,z) = \frac{z^A}{A},
    \end{equation}
     which enters Eq. (\ref{initCondz}).
  The generating function therefore reads
    \begin{equation}\label{solGenerA}
    \gener(t,z) = N( t ) + \frac{1}{2}\left( r  - \sqrt{r^2 + 4r(1-z)}\right) + \frac{ \sqrt{r^2 + 4r(1-z)} e^{-  \sqrt{r^2 + 4r(1-z)} t}}{ -1 + \frac{2 \sqrt{r^2 + 4r(1-z)}}{\frac{2}{A}(z^A-1) - r +  \sqrt{r^2 + 4r(1-z)}}   + e^{ -  \sqrt{r^2 + 4r(1-z)}   t}}.
   \end{equation}

  The second moment $M_2(t)$ of the family of densities $(c_k(t))_{k\geq 1}$ gives an order of magnitude
   of the square of the typical mass of the aggregates at time $t$. Using the exact expression of the generating function $\gener(t,z)$, 
    we can obtain this second moment from a Taylor expansion around $z=1$ (using $\sum_{k\geq 1} k c_k(t) = 1$ from the monomer-only boundary condition):
    \begin{equation}
     M_2(t) = \sum_{k\geq 1} k^2 c_k(t) = 1 + \frac{\partial^2\gener}{\partial z^2}(t,1).
    \end{equation}
    With the notations
    \begin{equation}
    \begin{split}
     \xi( z ) &:= \sqrt{r^2 + 4r(1-z)},\\
     \tau( z ) &:=   \frac{2}{A}(z^A - 1) - r +  \xi(z),
     \end{split}
    \end{equation}
  the generating function reads for initial conditions consisting of polymers of size $A$:
    \begin{equation}\label{solGenerM}
    \gener(t,z) = N( t ) + \frac{1}{2}\left( r  - \xi(z)\right) + \frac{ \xi( z) \tau( z )     }{ \left( - \tau( z ) + \xi(z) \right) e^{ \xi(z) t}  + \tau(z)}.
   \end{equation}
     We have the  following expansions around $z=1$:
   \begin{equation}\label{DL}
   \begin{split}
   \xi( 1-h ) &=  r\left( 1 + \frac{2}{r}h -\frac{2}{r^2}h^2 + o(h^2) \right)\\
    \tau( 1 - h) &=    (A-1)h^2  -\frac{2}{r}h^2 + o(h^2) = h^2 \left( -\frac{2}{r} + A - 1 \right)+ o(h^2).
     \end{split}
   \end{equation}
   The numerator in the last term of Eq. (\ref{solGenerM}) is therefore $O(h^2)$, and equivalent to $\xi(1)\tau(1-h)$ at small $h$. 
    The denominator
    is equivalent to $\xi(1)\exp( \xi(1)t)$, with $\xi(1)=r$.
    The term of order $h^2$  in the expansion of  Eq. (\ref{solGenerM})  yields:
 \begin{equation}
 \frac{1}{2}\frac{\partial^2\gener}{\partial z^2}(t,1) = \frac{1}{r} +\left(  -\frac{2}{r} + A- 1 \right)
   e^{- r t}.
 \end{equation}
 The second moment thefore reads
  \begin{equation}
     M_2(t) = \sum_{k\geq 1} k^2 c_k(t) = 1 + \frac{2}{r} + 2 \left(  -\frac{2}{r}+ A- 1 \right)  e^{-r t}.
    \end{equation}
The typical size of the aggregates converges expoentially  to the 
  steady-state value, at the resetting rate. Moreover, if the resetting rate is set to the special value
   \begin{equation}
    r_A := \frac{2}{A - 1},
   \end{equation}
   the second moment is kept constant (at $A$).

 \section{Discussion}
 
 In this paper we have obtained the generating function of the aggregation model with constant kernel subjected to resetting 
  at a constant rate (in the sense that aggregates of any size explode into monomers at independent Poisson-distributed times). We have  solved
  the master equation instead of relying on renewal equations. 
   This approach is natural in many-body systems whose  constituents are reset independently, and has already been used for local resetting 
    in \cite{miron2020diffusion,pelizzola2020simple}.\\
  
   The steady state of the model  is independent of the initial conditions and contains aggregates of all 
    sizes, whose  average density is a decreasing function of the size of the aggregate. 
  Moreover, this density assumes a scaling form, in which the size of the aggregate is rescaled according to the resetting rate.
   In the absence of resetting, the aggregation process is irreversible and the concentration of any aggregate of fixed size goes to zero,
    but the typical size of the aggregates grows systematically with time, and scaling occurs because  a change in time scale 
     preserves the mass distribution (provided the mass is rescaled by a time-dependent factor). When the model is subjected to resetting, 
      the large-time limit of the distribution of concentration is invariant if the size is rescaled  by a rate-dependent factor. Indeed low rates of resetting probe the large-time and large-size behaviour of the aggregation process.\\

   The density of aggregates of fixed size in the non-equilibrium steady state
   is maximised if  the resetting rate equals the inverse of the size (multiplied by the rate 
    of the aggregation process). The optimal resetting rate is higher for aggregates of low size, and of the same order of magnitude as the 
     rate of the aggregation process (which is set to $2$ in our calculations by picking the unit of time). The assumption of good mixing 
      is therefore valid for in model under resetting at optimal values, as long as it is valid for the  model without resetting.
   In the large-size  limit, the optimal resetting rate goes to zero, which is intuitive as a lower resetting rate is more favourable to large aggregates.\\
   
   Moreover, the generating function has been used to compute the second moment of the densities as a function of time. 
     For initial conditions consisting of polymers of fixed size (larger than $1$) and generic values of the resetting rate,
       the second moment goes exponentially to the steady state, and the rate of convergence is equal to the resetting rate.
          However, the second moment can be  constant, for a unique value of the resetting rate.\\

%depolymerisation
%The present model can be considered an extreme variant of depolymerisation models.\\

% extensions
 The constant kernel provides a workbench for modelling aggregation, as its simplicity allows to display remarkable properties 
  of the phenomenon, such as scaling. We have seen that it serves the same purpose when subjected to resetting. 
 Models of aggregation with size-dependent kernels, such as the sum and product kernels, have been proposed and solved \cite{melzak1953effect,scott1968analytic,mcleod1962recurrence}. It would be interesting 
  to subject them to resetting. Moreover, the resetting prescription itself could be generalised to become size-dependent.

\bibliography{bibRefsNew} 
\bibliographystyle{ieeetr}

\end{document}